\documentclass[pdflatex,sn-mathphys-ay]{sn-jnl}

\usepackage{url,hyperref,lineno,microtype,subcaption}
\usepackage[onehalfspacing]{setspace}
\usepackage{booktabs}
\usepackage{graphicx}
\usepackage{ltablex}
\keepXColumns 
\usepackage{tabularx}

\usepackage{longtable}
\usepackage{calc}
\usepackage{booktabs}
\usepackage{float}

\usepackage{geometry}
\begin{document}

\title[Article Title]{Recent advances in deep learning and language models for studying the microbiome} 


\author[1]{\fnm{Binhao} \sur{Yan}}
\equalcont{These two authors contributed equally to this work.}
\author[2]{\fnm{Yunbi} \sur{Nam}}
\equalcont{These two authors contributed equally to this work.}
\author[3]{\fnm{Lingyao} \sur{Li}}
\author[4]{\fnm{Rebecca A.} \sur{Deek}}

\author*[1]{\fnm{Hongzhe} \sur{Li}}\email{hongzhe@upenn.edu}
\author*[2]{\fnm{Siyuan} \sur{Ma}}\email{siyuan.ma@vumc.org}

\affil[1]{\orgdiv{Department of Biostatistics, Epidemiology, and Informatics, Perelman School of Medicine}, \orgname{University of Pennsylvania}, \orgaddress{\city{Philadelphia}, \state{PA}, \country{USA}}}
\affil[2]{\orgdiv{Department of Biostatistics}, \orgname{Vanderbilt University Medical Center}, \orgaddress{\city{Nashville}, \state{TN}, \country{USA}}}
\affil[3]{\orgdiv{School of Information}, \orgname{University of South Florida}, \orgaddress{\city{Tampa}, \state{FL}, \country{USA}}}
\affil[4]{\orgdiv{Department of Biostatistics and Health Data Science}, \orgname{University of Pittsburgh}, \orgaddress{\city{Pittsburgh}, \state{PA}, \country{USA}}}


\abstract{Recent advancements in deep learning, particularly large language models (LLMs), made significant impact on how researchers study microbiome and metagenomics data. Microbial protein and genomic sequences, like natural languages, form a \textit{language of life}, enabling the adoption of LLMs to extract useful insights from complex microbial ecologies. In this paper, we review applications of deep learning and language models in analyzing microbiome and metagenomics data. We focus on problem formulations, necessary datasets, and the integration of language modeling techniques. We provide an extensive overview of protein/genomic language modeling and their contributions to microbiome studies. We also discuss applications such as novel viromics language modeling, biosynthetic gene cluster prediction, and knowledge integration for metagenomics studies.}


\maketitle


\section{Introduction}


The study of microbiomes and metagenomics has significantly advanced our understanding of microbial communities and their complex interactions within hosts and environments. The microbiome refers to the collective genomes of microorganisms residing in a specific habitat, such as human body sites (e.g., gut, skin, airway) and environments (e.g., air, soil, water). Metagenomics research involves the direct profiling and analysis of these microbial communities' genomic sequences, bypassing the need for isolating and culturing individual members. This approach allows for a comprehensive assessment of microbial diversity, functions, and dynamics within their natural contexts.

The complex dependency encoded in metagenomic sequences represents gene/protein-, organism-, and community-level biological structures and functions. Examples include residue-residue contact patterns for protein 3D structures, functional relationship between genes and their regulatory, non-coding counterparts (e.g., promoters, enhancers), mobility for horizontal gene transfers, and genome-scale organization of functional modules (e.g., operons and biosynthetic gene clusters). Such dependency patterns, when interrogated at the revolutionary scale (i.e., encompassing many diverse organisms and environments), can capture fundamental biological properties as shaped over time by evolutionary processes, thus representing a meaningful ``language of life''. On the other hand, the availability of microbial genomic sequences, both annotated and unannotated for biological properties (e.g., UniRef, \cite{suzek2007uniref}; MGnify, \cite{richardson2023mgnify}), drastically increased over the past decade due to advancements in next-generation sequencing protocols, bioinformatics, and computational capacities. The availability of these metagenomic ``big data'' suggests that, given capable modeling architecture and capacity, microbiomes' evolutionary and functional dependency structures can be computationally learned, represented, and utilized for studying the microbiome. 

To this end, advances in powerful artificial intelligence (AI) methods regarding the design and training of highly complex, large-scale deep learning models have been adopted to characterize microbial genes and genomes from large-scale metagenomic sequences, offering powerful tools for extracting, interpreting, and integrating complex microbiome data \citep{hernandez2022machine}. In particular, inspired by the recent breakthrough of large language models (LLMs) in dealing with natural language tasks, similar methods have been developed and applied for modeling protein and genomic languages of life. To avoid conflating nomenclature, we reserve ``LLM'' in this review exclusively for large language models (e.g., ChatGPT \citep{liu2023summary}) and instead use terms such as ``protein language model'' and ``DNA language model'' to more explicitly refer to genomic sequence models. Indeed, whereby natural languages are organized in sequential words and phrases which form the basic units of modeling (``tokens''), microbial genomic elements are similarly organized as sequences of nucleotide base pairs (for genomic DNA) or amino acids (AA, for proteins). Given the complexity of genomic dependency structures, metagenomic research was fast to adopt advanced language modeling techniques for studying microbial community sequences, with models spanning different genomic scales (microbial proteins, contigs, genomes, and communities) and designed for a variety of tasks (Figure \ref{fig1}), yielding promising performance improvement and novel applications.

This review aims to provide a survey of recent developments in deep learning and language modeling for analyzing microbiome and metagenomics data. We focus on problem formulation, the datasets required to address these questions, and how sequence-based language models are integrated into deep learning algorithms. In Section 2, we briefly discuss the typical language model architecture from recent LMM breakthroughs and how they can be applied to genomic sequence modeling. We discuss in Section 3 two broad classes of language models for microbiome studies, namely, \textit{protein} language models and \textit{DNA/genomic} language models, distinguished by their drastically different range of genomic ``contexts'' for sequence dependency structures. We then review three specific applications of high interest to the field in Sections 4-6, namely, novel viromics language modeling, models for predicting biosynthetic gene clusters (BGCs), and knowledge integration in metagenomics studies aided by natural language LLMs. We conclude with prospective remarks and discussions in Section 7.

\section{Brief review of LLMs and their extension towards modeling the language of life}
LLMs are advanced foundation models specifically designed to understand and generate human language. They can perform a wide range of natural language processing (NLP) tasks, such as question-answering, information extraction, and text summarization \citep{chang2024survey, li2024scoping}. The scalability, versatility, and contextual understanding of LLMs can be attributed to two key factors. First, LLMs are trained on massive datasets that encompass diverse linguistic patterns, enabling them to learn complexities and nuances in languages as sequences of tokens (i.e., words and phrases). Second, LLMs are built on the transformer architecture, which consists of an encoder and a decoder and uses self-attention mechanisms to process input sequences. The attention mechanism efficiently encodes dependency structures of sequential tokens, vastly increases the learnable lengths of long-range dependencies, and encodes tokens and sequences accounting for their upstream and downstream neighboring ``contexts''. This allows for efficient processing of sequential data, enabling LLMs to provide a meaningful representation of input text and generate coherent and contextually relevant output text based on input prompts \citep{vaswani2017attention}.

\begin{figure}
\includegraphics[width=0.95\textwidth]{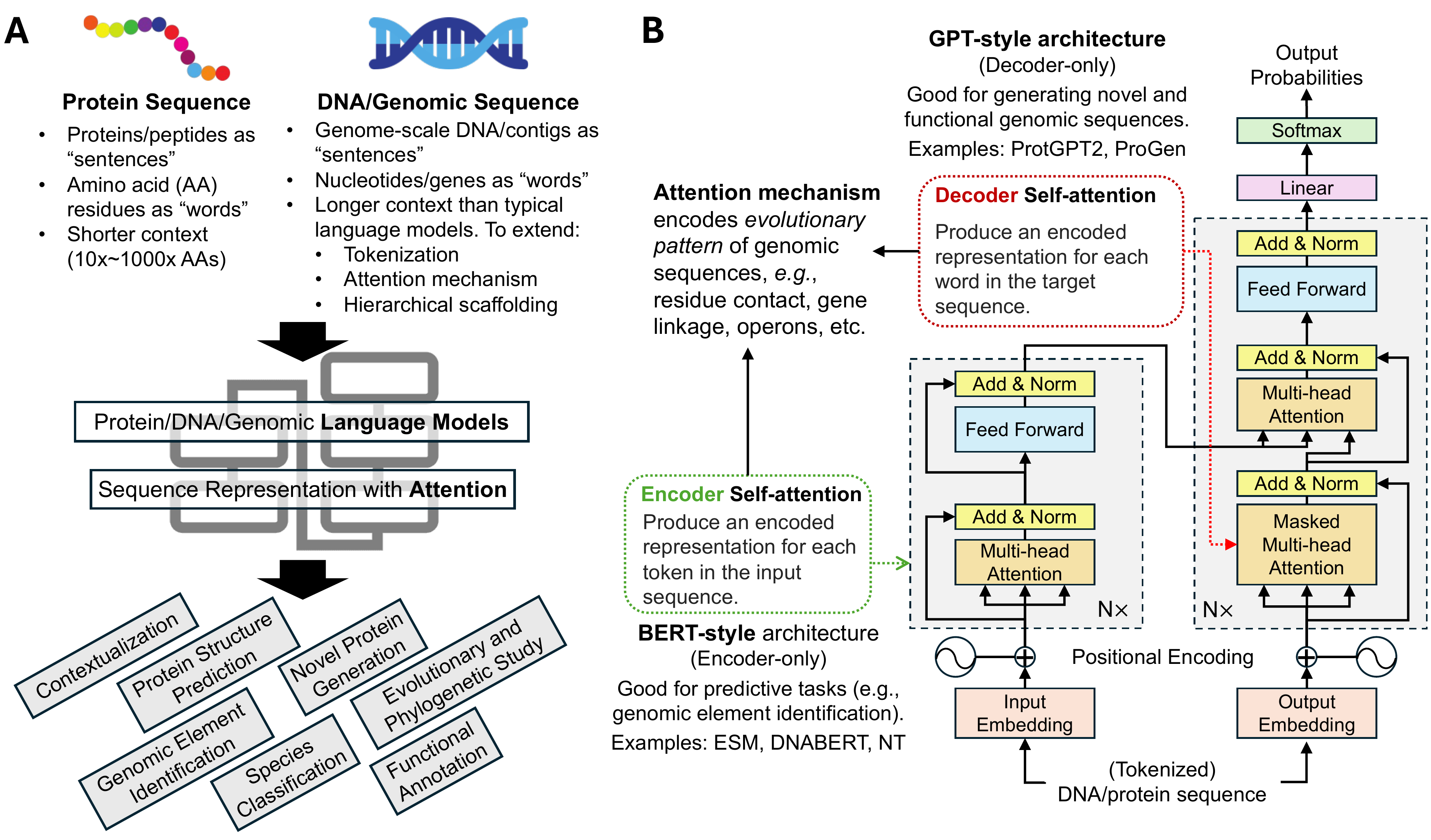}
\caption{Review of protein/DNA/genomic language models as applied to metagenomic studies. A. Protein and genomic sequences share similar properties as natural language sequences, with amino acids or nucleotides as units of sequences (``tokens''). The complex dependency structure of protein/gene-level or genomic-scale sequences can then be modeled by language model techniques, such as the transformer-based attention mechanism for various downstream tasks. B. Review of encoder- and decoder-style transformer attention mechanisms and their applications in metagenomic studies. Decoder-style model architecture (similar to that of BERT) aims to provide a meaningful representation of genomic sequences and is useful for downstream predictive tasks. Encoder-style model architecture (similar to that of ChatGPT) generates new sequences given past tokens and is most useful for generative tasks such as novel protein design.}\label{fig1}
\end{figure}

Inspired by LLMs, language models in microbiome research often employ a similar architectural design (Figure \ref{fig1}). These language models of genomic sequences \citep{ligeti2024prokbert, shang2023phatyp, mardikoraem2023generative} thus provide an improved representation of sequences with richer context and can be scaled up to and trained at impressive complexities (up to billions of model parameters). Such models often include a transformer encoder component that processes input sequences---such as protein or DNA sequences---and converts them into high-dimensional representations that capture essential features of input sequences in their contexts. The attention mechanism in these models assigns different weights to various parts of the sequence, characterizing dependency structures shaped over evolution and allowing the model to focus on relevant regions. This focus ensures that the model prioritizes areas within a sequence that are most significant for biological interpretation. Similar to the application of BERT for various natural language tasks \citep{devlin2018bert}, the encoded representation can then be used for tasks such as contextualizing microbial genes and genomic sequences in their broader genomic neighborhoods \citep{hwang2024genomic}, predicting the structure and functions of protein given their sequences \citep{lin2023evolutionary}, and segmentation and identification of specific regulatory elements across microbial genomes \citep{Zhou2025DNABERT2}. In comparison, decoder-style models focus on generating output sequences, given the encoded representation of past sequence tokens. This is more similar to GPT-style LLMs \citep{brown2020language},  whereby the task towards microbiome application often involves the generation of new, functional, and viable protein sequences \citep{ferruz2022protgpt2,madani2023large, jin2024prollm}.

\section{Language modeling of proteins, contigs, and genomes of the microbiome}


To facilitate the survey, we categorize existing language models for metagenomic sequences into two classes: (1) models on the protein/gene scale and (2) those on the genome scale. The first, which we term \textit{protein} language models (Table \ref{tab:PLMs}), fits well within the context length for transformers since microbial proteins are generally under 1,000 AAs (tokens). In contrast, \textit{DNA} or \textit{genomic} language models (Table \ref{tab:GLMs}) often require additional techniques to extend their operating ranges due to the large scale of microbial contigs or whole genomes. For example, the bacterial genome typically ranges from 0.5 to 10 million base pairs, a scale that often far exceeds the context window of transformers. In addition, the two classes target different applications: protein language models are used for designing and predicting individual proteins, while DNA/genomic language models examine genes and proteins within their broader genomic contexts as well as intergenic regions.

\begin{table}[h!]
    \centering
    \caption{Protein Language Models.}
    \begin{tabular}
    {p{0.1\textwidth}p{0.2\textwidth}p{0.15\textwidth}p{0.2\textwidth}p{0.2\textwidth}}
    \\ \toprule
    \textbf{Model} & \textbf{Model Architecture} & \textbf{Usage} & \textbf{Relevance to the Microbiome} & \textbf{Additional Notes}  \\ \midrule
        Antimicrobial peptide (AMP) prediction \citep{ma2022identification} & 
        The best combination of LSTM, attention-based and BERT models. & 
        Identifying candidate AMPs from human microbiome data that were further validated experimentally. &
        Training/validation data and application focus on bacterial peptides.
        & N/A \\ \hline
    
        ProtGPT2 \citep{ferruz2022protgpt2} &
        Transformer decoder model that matches that of GPT2. &
        Generating novel proteins. &
        Universal training and validation sequences include microbial proteins. &
        Byte Pair Encoding (BPE) tokenization improves model performance. \\ \hline    
        
        ProGen \citep{madani2023large}; \newline Progen2 \citep{nijkamp2023progen2} & 
        Standard transformer decoder with left-to-right causal masking. & 
        Generating viable and novel proteins with controlled functions. & 
        Universal training and validation sequences includes microbial proteins. \newline 
        Generated viable and experimentally validated antibacterial proteins. & 
        Model size study suggests even huge models ($>$6 billion parameters) are far from overfitting. \\ \hline  
        
        ESM-1b \citep{rives2021biological}; \newline
        ESM-2, ESMFold \citep{lin2023evolutionary} &
        ESM-1b/ESM-2: BERT-like masked token architecture; \newline
        ESMFold: Folding NN (based on ESM-2 representations) composed of folding blocks + structure prediction module. & 
        ESM-1b/ESM-2 provides meaningful sequence representations. \newline
        ESMFold provides fast and accurate prediction of protein 3D structure based on sequences. &
        Universal training and validation sequences include microbial proteins. \newline 
        Provides database with prediction of large number of metagenomic protein sequences. & 
        Protein sequence attention pattern can predict structural residue contact probability in a zero-shot fashion. \newline
        Language model architecture allows faster structure prediction compared to SOTA models based on multiple sequence alignment. \\
        \bottomrule
    \end{tabular}
    
    \label{tab:PLMs}
\end{table}

\subsection{Protein language models for novel protein generation}

Existing protein language models applied towards microbiome studies are summarized in Table \ref{tab:PLMs}. We highlight two specific applications, namely, the generation of novel proteins and the prediction of their functions and structures. The dependency structure of amino acids across known microbial proteins is learned and utilized to generate artificial, potentially novel protein sequences by protein language models such as ProGen \citep{madani2023large} and ProtGPT2 \citep{ferruz2022protgpt2}. This is performed in an autoregressive fashion, often with decoder-only architecture similar to that of the GPT language models, whereby the likely AA at the next position is predicted given the sequence of preceding residues. If trained across a sufficiently large variation of raw occurring microbial protein spaces (millions or more protein sequences), models with enough flexibility can learn the inherent evolutionary patterns that natural protein sequences harbor and thus generate artificial proteins that are functionally viable like natural proteins. 
 
 To this end, ProtGPT2 was based on the GPT-2 architecture and trained on 50 million sequences spanning the entire protein space. Proteins generated by the model in return displayed propensities of amino acid sequences akin to those of natural proteins, but can still cover under-explored protein sequence regions. ProGen and its iteration \citep{nijkamp2023progen2} performed similar modeling tasks, and additionally (1) allowed the inclusion of ``tags'' to specify protein properties for generating proteins in a more controllable fashion, and (2) experimentally verified that model-generated de novo protein sequences were sufficiently distinct from natural proteins but demonstrated functional viability comparable to them. Of note, while these models were typically trained to cover the universal protein space (e.g., UniRef-50), both models highlight good coverage of microbial protein properties. ProGen specifically validated the antibacterial functional property of its generated novel proteins that were comparable to natural lysozyme.

\subsection{Protein language models for function and structure prediction}
 
 Related to, but different from the task of generating novel protein sequences, prediction-focused protein language models are primarily concerned with predicting proteins' biological properties (e.g., 3D structures, functions) based on their AA residue sequences. Encoder-style language model architectures such as that of BERT are of particular relevance, as these models aim to learn the best representation of each token (i.e., AA) given the broader sequence context and thus can represent entire sequences in a meaningful, efficient manner.  For example, \cite{Elnaggar2022ProtTrans} developed several LMs for protein sequences, including two auto-regressive models (Transformer-XL, XLNet) and four auto-encoder models (BERT, Albert, Electra, T5) on data from UniRef and BFD containing up to 393 billion amino acids. Transformer Uniref90 MT from the Protein BERT project can be downloaded from the project GitHub repository (https://github.com/nadavbra/protein\_bert) and protein sequences are embedded using the function in the protein Bert python package. Such representations can then be fed as the input to downstream predictive models, often also realized with neural networks (NN), for various tasks.
 
 For predicting protein structures, with scaling language models from 8 million to 15 billion parameters, the ESM-2 model \citep{lin2023evolutionary} effectively internalizes evolutionary patterns directly from protein sequences. The learned attention patterns provided a low-resolution protein structure, corresponding to residue-residue contact maps. This was further combined with a downstream predictive module to form the ESMFold model, which offers direct inference from sequence to protein 3D structures and achieved comparable performances as SOTA protein structure prediction models (e.g., AlphaFold2). Of relevance to the microbiome, the authors applied their model to construct an atlas of predicted structures of over 600 million metagenomic protein sequences. Another group of predictive tasks aims to mine biological functions based on protein sequences. As a representative, \cite{ma2022identification} focused on predicting antimicrobial peptides (AMPs) as products of the gut microbiome. They constructed the best combination over several language models, including one of the BERT architecture, to computationally mine AMP candidates from gut metagenomic studies. With additional computational filtering and experimental validation, they demonstrated that identified candidates were effective against multi-drug-resistant bacteria and demonstrated microbial membrane disruption in mechanistic studies. Such studies represent the potential of language model-aided computational efforts toward human and environmental microbiome studies for high-throughput mining of microbial structural and functional properties.

\subsection{DNA language models at the genomic scale}

The full review of DNA/genomic language models is provided in Table \ref{tab:GLMs}. As discussed above, microbial genomes have drastically increased scales compared to single genes or proteins. Genomic sequences also possess much sparser biological information than proteins, containing intergenic regions with both functional and junk DNA elements. The DNA sequence vocabulary also only consists of four different types of nucleotides, less than the 20 different AAs that typically constitute protein sequences. As such, language models that operate on the genome scale require additional considerations than protein models and can be further divided into two categories. The first type, often termed in literature as DNA language models, focuses on modeling DNA sequences truly on the full genome scale of organisms, e.g., DNABERT \citep{Zhou2025DNABERT2}, Nucleotide Transformer (NT, \cite{dalla2023nucleotide}). As such, they adopt techniques such as specialized tokenization, alternative attention patterns, and hierarchical modeling architecture to drastically extend model contextual lengths. An important advantage of this approach is that it allows for the representation and identification of non-coding functional elements on the DNA (e.g., promoters).

\begin{footnotesize}
\begin{longtable}[hc]
    {p{0.12\textwidth-2\tabcolsep}
    p{0.22\textwidth-2\tabcolsep}
    p{0.22\textwidth-2\tabcolsep}
    p{0.19\textwidth-2\tabcolsep}
    p{0.25\textwidth-2\tabcolsep}} 

\caption{DNA/Genomic Language Models.}
\label{tab:GLMs}

\\ \toprule
\textbf{Model} & \textbf{Model Architecture} & \textbf{Usage} & \textbf{Relevance to the Microbiome} & \textbf{Additional Notes}
\\ \hline 
DNABERT \citep{ji2021dnabert};\newline
    DNABERT-2 \citep{Zhou2025DNABERT2}
    & 
    Encoder-only transformer architecture with masked modeling. \newline DNABERT-2 incorporates new techniques such as Attention with Linear Biases to increase context length and Flash Attention to increase computation and memory efficiency & 
    Human (DNABERT) and multi-species (DNABERT-2) for representing genomic nucleotide sequences applicable for downstream tasks.  \newline
    Demonstrated utilities in various tasks: genomic element prediction (promoter, enhancer, transcription factor, epigenetic marks), microbial species classification &
    DNABERT-2 training/validation based on multi-species genomes including bacteria and fungi. \newline
    Evaluated for microbial gnomic element prediction and species classification. & 
    Also provides system of benchmarking tasks for evaluating DNA language models. \newline
    BPE tokenization improves model performance.
    \\ \hline

    gLM \citep{hwang2024genomic} & 
    RoBERTa-based transformer architecture.\newline
    Genes (embedding from ESM-2) are tokens and microbial contigs (15-30 genes) are sequences. & 
    Enriching microbial genes' representations with longer-range genomic context. Providing contextualized gene function prediction and characterizing higher-order genomic features. &
    Training/validation data and application focus on microbial genomes. & 
    ``Contextualization'' involves learning (transformer-based) representations microbial genes' longer-range genomic contexts. Such representations encode enriched genomic information such as mobility for horizontal gene transfer and operon membership.
    \\ \hline

    NT \citep{dalla2023nucleotide};\newline
    SegmentNT \citep{de2024segmentnt}
    & 
    NT: encoder-only transformer architecture with masked modeling. \newline 
    SegmentNT: a segmentation NN head based on NT embedding & 
    NT provides human-focused and multi-species (DNABERT-2) representation for genomic nucleotide sequences applicable for downstream tasks. \newline
    SegmentNT specializes in predicting genomic elements based NT representation. &
    NT has multi-species version incorporating bacterial and fungal genomes. \newline
    SegmentNT does not incorporate microbial genomes. & 
    Performance of multi-species model was demonstrated to match or outperform human-only model on tasks specific for human genomes.
    \\ \hline
    
    Species-aware DNA language models \citep{karollus2024species} & Standard DNABERT architecture was adopted. & (a) to learn meaningful species-specific and shared regulatory features across evolution (b) to transfer these features to unseen species. & Trained on non-coding regions from $>$800 fungal species spanning over 500 million years of evolution. & Focus on non-coding DNA and regulatory elements. \\ \hline
    FGBERT \citep{duan2024fgbert} & Joint objectives of (a) masked gene modeling with a context-aware tokenizer and (b) contrastive learning with data augmentation and negative sampling to capture the functional relationships between genes. & Downstream tasks include gene operons, functional genes, genome pathogenes, and nitrogen cycle prediction. & Pre-trained on 100 million metagenomic sequences. & First metagenomic pre-trained model encoding (a) context-aware and (b) function-relevant representations of metagenomic sequences. Protein-based gene representations converted from the DNA sequence from metagenomic sequences, to protein sequence using ENA, and then to ESM-2 representations. \\ \hline
    ProkBERT family \citep{ligeti2024prokbert} & Encoder-only masked language modeling with the newly introduced Local Context-Aware (LCA) tokenization. & Generate nucleotide sequence representation. Applied downstream tasks include (a) bacterial promoter prediction and (b) bacteriophage identification. & Bacteriophages have a significant role in the microbiome, influencing host dynamics and serving as essential agents for horizontal gene transfer. & The implementation of masked language modeling (MLM) with LCA requires slight variations in masking tokens: to prevent trivial restoration from locality, the model needs to ensure neighboring tokens to be masked as well. \\
    \bottomrule
\end{longtable}
\end{footnotesize}

Tasked with the ambitious goal of providing a generic, ``foundation'' model for genomes, models such as the DNABERT and NT aim to provide meaningful, contextualized representations of genome-scale DNA sequences that can be used to predict their functional properties and molecular phenotypes. Trained on genomes spanning hundreds of organisms (including genomes from microbial species) and based on encoder-style model architectures, these models are then utilized towards tasks such as predicting genomic elements (promoter, enhancer, transcription factor, epigenetic marks) and differentiating microbial species. While studies of human genomes are still the focus, they do demonstrate transferrable of learned representations across species to metagenomes, as well as improved model performance when the combination of diverse genomes was included during model training. On more reduced scales, microbial DNA language models are focused on learning the genomic pattern of specific organisms and specific genomic elements. \cite{karollus2024species} for example trained a DNABERT-like model specifically for non-coding regions up- and down-stream of gene sequences from fungal species, and demonstrated that the learned sequence representations can capture motifs and regulatory properties of these elements, in contrast to the background and non-coding sequences. While current DNA language models are still limited in training data and model capacity (the NT at its largest scale was trained with 2.5 billion parameters on 850 species) to truly operate as the foundational representation of diverse genomes, we anticipate significant progress in the near future aided by rapid development on language model scales and computational power.

\subsection{Genomic language models contextualize genes and gene clusters.} 

Alternatively, another group of metagenome language models examines medium- to long-range contexts between genes, often operating on the contig scale and excluding intergenic sequences. We term these as genomic models as an intermediate approach between protein and DNA language models. These models often adopt hierarchical scaffolding across genes (genes themselves are embedded by protein language models), to provide a contextualized and richer representation of genes in their broader genomic neighborhood. Gene properties such as their differential functions across microbes and genome/community-scale organization (horizontal gene transfer, operon membership) can then be further interrogated, which is not possible in protein language models where they are modeled in isolation from each other.

In comparison to full-scale DNA language models, genomic language models such as the gLM \citep{hwang2024genomic} and FGBERT \citep{duan2024fgbert} instead focus on contig- to genome-scale organization of microbial genes (see Table \ref{tab:GLMs}). gLM, for example, adopts EMS-2 protein embeddings for each gene and models their genomic dependency structures on the contig (15 to 30 genes in length) scale. This enables, first, the enrichment of each gene's embedding in its broader genomic texts. Genes' ``higher-order'', genome- and community-level functional properties can be further delineated that are indistinguishable from protein-scale language modeling alone, such as differential gene functions in different biomes and microbial species, as well as their self-mobility in horizontal gene transfer events across genomes. Secondly, the organization of gene clusters in linkage with each other on the genome can also be represented, whereby subsets of model attention patterns from gLM and FGBERT both demonstrated correspondence with operon memberships. The longer-scale organization of biosynthetic gene clusters is also relevant and discussed in a dedicated section as a specialized task. As population-scale studies of the microbiome often focus on gene- or pathway-level sample profiles, such genomic language models provide practical intermediate solutions to enrich microbiome studies using recent language model advancement with microbial gene elements' broader genomic contexts.

\section{Language models for virome annotation and  virome-host  interactions}

The human virome consists of viruses that infect eukaryotic cells (eukaryotic viruses) and prokaryotic viruses, also known as bacteriophages. The gut virome is a vital component of the microbiome in the human gut, consisting mainly of viruses that infect bacteria (bacteriophages or phages), along with other viral species that may infect eukaryotic cells. The virome plays a crucial role in maintaining gut health by influencing the bacterial population dynamics, shaping immune responses, and potentially affecting the overall metabolic environment of the gut.  

Metagenomic sequencing of the gut microbiome provides a wealth of information that aids in identifying viruses, especially bacteriophages, which are key players in viral-bacterial interactions. One important method for studying these interactions is through CRISPR spacers, which serve as a molecular record of past viral infections in bacterial genomes. CRISPR-Cas systems are a bacterial immune defense mechanism that targets invading viruses, primarily bacteriophages \citep{dion2021streamlining}. There has been significant interest in applying recently developed protein or DNA sequence language models in virome sequence identification and annotation, as well as in building predictive models for virus-bacterium interactions based on sequence data.

\subsection{Virome sequence annotation and identification}

Annotation of viral genomes in metagenomic samples is a crucial first step in understanding viral diversity and function. Current annotation approaches primarily rely on sequence homology methods, such as profile Hidden Markov Model (pHMM)-based approaches. However, these methods are limited by the scarcity of characterized viral proteins and the significant divergence among viral sequences. To address these challenges, \cite{flamholz2024large} applied curated virome protein family (VPF) databases alongside recently developed protein language models (PLMs). They demonstrated that PLM-based representations of viral protein sequences can capture functional homology beyond the reach of traditional sequence homology methods.
Their reference annotations were derived from the Prokaryotic Virus Remote Homologous Groups (PHROGs) database, a curated library of VPFs designed to detect remote sequence homology. PHROGs is manually annotated into high-level functional categories and contains 868,340 protein sequences clustered into 38,880 families, of which 5,088 are assigned to 9 functional classes. Using these data, \cite{flamholz2024large} showed that PLM-based representations of viral proteins can effectively predict their functions, even in the absence of close sequence homologs.


\cite{peng2024viralm} developed a viral language model (ViraLM)  that adapts the genome foundation model DNABERT-2 \citep{Zhou2025DNABERT2} for virus detection by fine-tuning the model for a binary classification of novel viral contigs in metagenomic data. 
  DNABERT-2 has been pre-trained on a vast array of organisms, acquiring valuable representations of DNA sequences, which is particularly useful for distinguishing viral sequences from those of other species.
To adapt the genome foundation model for virus detection, they fine-tuned this model for a binary classification task with two labels: viral sequences vs. others, where they constructed a substantial viral dataset comprising 49,929 high-quality viral genomes downloaded from the NCBI RefSeq, spanning diverse taxonomic groups as positive samples. The negative data (245,734 non-viral sequences) are complete assemblies of bacteria, archaea, fungi, and protozoa, also downloaded from the NCBI RefSeq. The genomes are randomly cut into short contigs ranging from 300 bp to 2k bp to minic variable-length contigs in the metagenomic data. They observed that the model initialized using the pre-trained foundation model converges faster and performs better in virus contig identification. 

\subsection{Deep learning and LLM methods for virome-host interaction}

One important problem in virome research is to predict which viruses can infect which hosts,  a crucial step for 
understanding how viruses interact with hosts and cause diseases. Virome-host interactions also play a crucial role in understanding and defining phage therapy, which uses bacteriophages (viruses that infect bacteria) to treat bacterial infections. Currently, there are no high-throughput experimental methods that can definitively assign a host to the uncultivated viruses. 

A number of computational approaches have been developed to predict unknown virus-host associations. The coevolution of a virus and its host left signals in their genomes, which have been exploited for computational prediction of virus-host associations. 
The alignment-based approaches search for homology such as prophage \citep{Roux2015VirSorter} or CRISPR-cas spacers \citep{Staals2013CRISPRCas,Horvath2010CRISPRCas}. Algorithms like BLAST (Basic Local Alignment Search Tool) are commonly used to align viral sequences with host genome sequences to detect homology. This can reveal conserved regions in viral and host proteins, such as receptor-binding domains that allow viruses to enter host cells. In contrast,  alignment-free methods use features such as k-mer composition, codon usage, or GC content to measure the similarity between viral and host sequences or to other viruses with a known host.   By identifying which viral genomes contain sequences matching a bacterium's CRISPR spacers, researchers can infer potential virus-host interactions. However, this approach is limited by the set of known CRISPR spacers.

As a comparison, predicting virus-host interactions based on k-mer matching and codon usage analysis is another powerful approach for identifying novel viral-bacterial interactions. Codon usage refers to the frequency with which different codons are used to encode amino acids in a genome. When a virus's codon usage matches that of its host, it suggests that the virus has evolved to efficiently exploit the host's translational machinery, enhancing its ability to replicate within that host. This is a critical factor in predicting potential virus-host interactions. By performing joint analysis of codon usage and other genomic features, researchers can achieve more accurate predictions regarding which host species are susceptible to particular viruses.

Since these genomic features are embedded in the viral or bacterial genomes, it is possible to learn these features automatically using machine learning and AI methods. \cite{Liu2023PredictionVirusHost} developed evoMIL for predicting virus-host association at the species level from viral sequence only. They used datasets that were collected from the Virus-Host database VHDB, (https://www.genome.jp/virushostdb/), which contains a manually curated set of known species-level virus-host associations collated from a variety of sources, including public databases such as RefSeq, GenBank, UniProt, and  ViralZone and evidence from the literature surveys \citep{Liu2023PredictionVirusHost}. For each known interaction, this database provides NCBI taxonomic ID for the virus and host and the Refseq IDs for the virus genomes.   The final data set includes  17,733 associations between 12,650  viruses and 3740 hosts that were used to construct binary datasets for both prokaryotic and eukaryotic hosts. For each of the hosts, an evoMIL model is built to predict the possible interacting viruses. 

They then applied the pre-trained ESM-1b model to transform protein sequences into fixed-length embedding vectors, which serve as features for downstream binary and multi-class classification. Additionally, they applied multiple instance learning (MIL) \citep{Maron1997MILFramework}, where multiple instances are grouped together with a single label, and are classified as a whole. They employed attention-based MIL \citep{Ilse2018AttentionMIL} for each host. Specifically, for each host, they collected the same number of positive and negative viruses and then obtained embeddings of protein sequences from viruses obtained by the pre-trained transformer model ESM-1b. To handle the input length of the PLMs, they split the protein sequences of viruses to sub-sequences for generating embeddings. An attention-based MIL was applied to train the model for each host dataset by protein feature matrices of viruses.  The resulting models can be used to predict whether a new virus interacts with a host for which a corresponding predictive model has been developed.

In addition to species-level virus-bacterium interaction prediction, \cite{Gaborieau2023.11.22.567924} introduced a novel dataset and prediction model that focuses on phage-bacteria interactions at the strain level, utilizing genomic data of 403 natural, phylogenetically diverse, \textit{Escherichia} strains and 96
bacteriophages. Their findings highlight that bacterial surface structures, such as lipopolysaccharides (LPS) and capsules, play a critical role in determining these interactions. Specifically, they identified bacterial surface polysaccharides as key adsorption factors that significantly enhance the accuracy of interaction predictions. This offers a valuable dataset for developing phage cocktails to combat emerging bacterial pathogens.

\section{Deep learning and language models for prediction of biosynthetic gene clusters}

Microbial secondary metabolites are chemical compounds that exhibit a broad range of functions and have great potential in pharmaceutical applications, such as antimicrobial agents and anticancer therapies. These bioactive small molecules are usually encoded by clusters of genes along the bacterial genome known as Biosynthetic Gene Clusters (BGCs). Although accurate, experimental validation of BGCs is laborious and costly. High-throughput sequencing techniques, alongside advanced genome assembly algorithms, have enabled people to access the vast amount of bacterial genomic data. The genomic sequence data serves as a rich resource for BGCs mining, allowing researchers to better understand the functional potential of bacteria and discover new secondary metabolites or natural products.

Machine learning-based algorithms have been developed for the detection of BGCs in microbial genomes. antiSMASH \citep{medema2011antismash} identifies candidate BGCs through multiple sequence alignment based on the profile hidden Markov model (pHMM) library constructed from experimentally characterized signature protein or protein domains, subsequently filtering these candidates using curated rules based on expert knowledge. PRISM \citep{skinnider2017prism} employs a similar approach by searching through an HMM library. ClusterFinder \citep{cimermancic2014insights} utilizes a hidden Markov-based probabilistic algorithm to identify known and unknown BGCs. Extending beyond these methods, MetaBGC \citep{sugimoto2019metagenomic} integrates segmented pHMM with clustering strategies, making it possible to detect BGCs directly from metagenomic reads.

Despite the success of existing machine learning-based algorithms, traditional machine learning models cannot handle the long-range dependencies between genome sequences and cannot transfer knowledge from other datasets, thereby resulting in a lower power of detecting the new BGCs. Several machine learning frameworks, including those with transformer-type language modeling architecture,  have been developed specifically for predicting bacterial BGCs. These models leverage advanced computational techniques to analyze genomic data and identify regions that encode for biosynthetic pathways. Many existing methods use sequences of the protein family domains (Pfams) to characterize the BGCs and bacterial genomics. Proteins are generally composed of one or more functional regions, commonly termed domains. Different combinations of domains give rise to the diverse range of proteins found in nature. The identification of domains that occur within proteins can therefore provide insights into their function.

\subsection{Deep learning methods for BGC prediction}
DeepBGC is a deep learning-based tool that uses a combination of convolutional neural networks (CNNs) and recurrent neural networks (RNNs) to predict and classify BGCs in bacterial genomes. It processes raw genomic sequences to identify BGCs and provides detailed annotations of their functional components \citep{hannigan2019deep}. e-DeepBGC further extended DeepBGC to incorporate functional description of protein family domains and to utilize the Pfam similarity database in data augmentation \citep{liu2022deep}. 
Pfam also generates higher-level groupings of related entries, known as clans. A clan is a collection of Pfam entries which are related by similarity of sequence, structure or profile-HMM.
\cite{rios2023deep} developed a deep learning model that leverages self-supervised learning to detect and classify BGCs in microbial genomes. This approach aims to improve the accuracy and efficiency of BGC identification and predict the types of natural products they produce.

\subsection{BGC prediction based on language models}
\cite{lai2023deciphering} introduced BGC-Prophet, a neural network model that leverages natural language processing (NLP) techniques to analyze genomic sequences as linguistic data, identifying patterns indicative of biosynthetic gene clusters (BGCs). This innovative approach enables the model to grasp the complex syntax and semantics inherent in genetic sequences. The input to BGC-Prophet consists of embeddings represented by 320-dimensional vectors, generated through ESM-2 \citep{lin2023evolutionary}. The model architecture integrates convolutional neural networks (CNNs) with transformer-based models, a hybrid design that effectively manages the sequential nature of DNA data, thereby enhancing the accuracy of BGC detection and classification. Table \ref{BGC.tbl} compares these methods, highlighting the deep learning models and primary data sources used for training.

\begin{table}[h]
\caption{Deep learning methods for BGC prediction}\label{BGC.tbl}
\centering
\begin{tabular}{ccccc}
\toprule
\textbf{Algorithm} & \textbf{Model} &\textbf{Pretraining}  & \textbf{Level} & \textbf{Primary source data}\\
\midrule
DeepBGC & BiLSTM & No & Pfam &  BGCs from \citep{cimermancic2014insights} + MIBiG\\
e-DeepBGC & BiLSTM & No & Pfam & MIBiG \cite{medema2015minimum}\\
BiGCARP & ByteNet & Yes & Pfam & antiSMASH \citep{blin2019antismash} + MIBiG \\
BGC-Prophet & Transformer & Yes & Amino acids & GTDB \cite{parks2022gtdb} + MIBiG \\
\bottomrule
\end{tabular}
\end{table}

\begin{figure}
\centering
\includegraphics[width=0.9\textwidth]{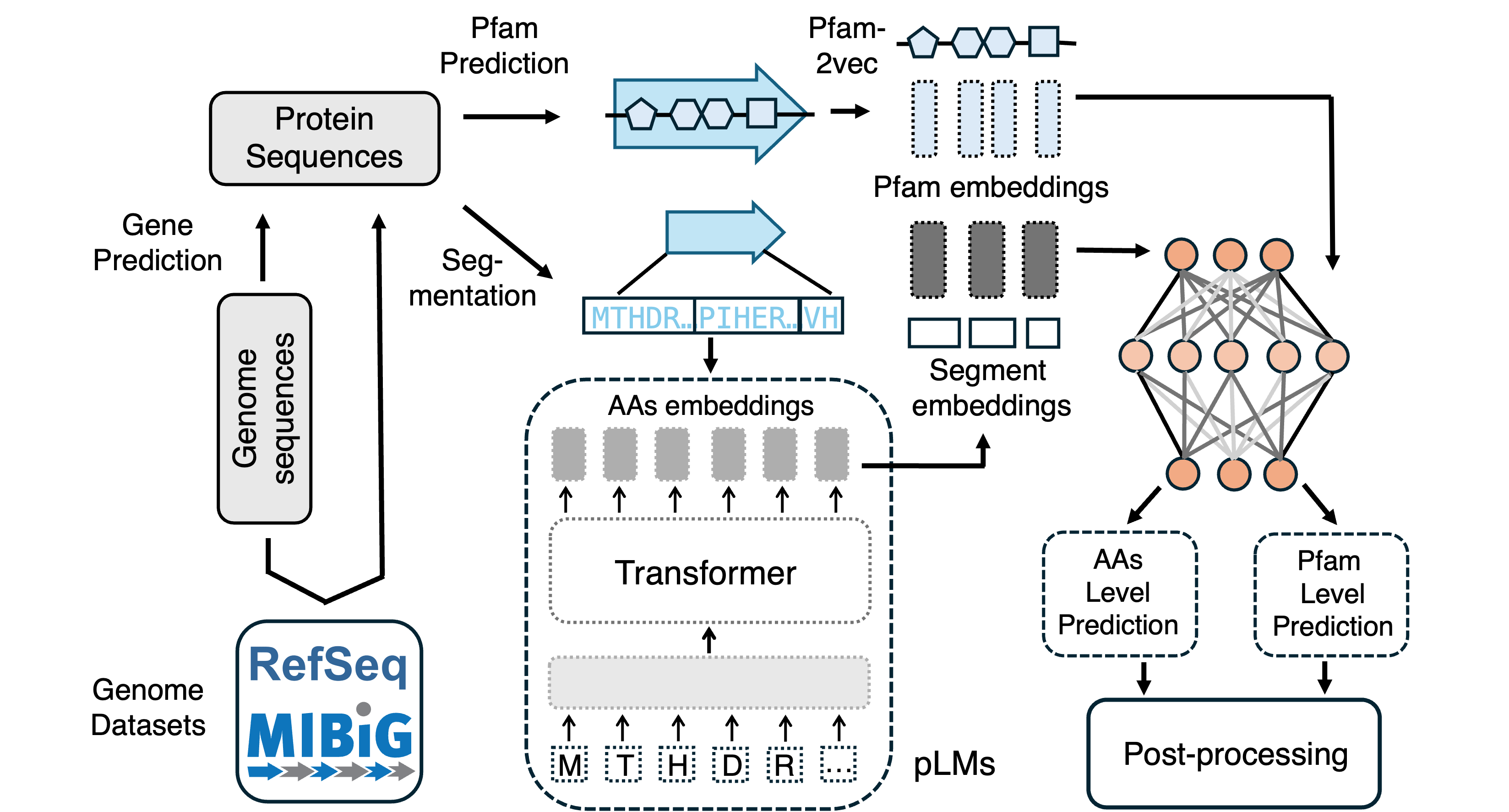}
\caption{A comparison of BGC prediction based on pfam2vec embedding for Pfam level prediction and embedding based on PLMs for amino acid level prediction.    }\label{AA-seq}
\end{figure}

Figure \ref{AA-seq} illustrates the differences in BGC prediction when performed at the Pfam level versus the amino acid level. Positive samples can be derived from segmenting amino acid sequences within biosynthetic gene clusters (BGCs) in the MIBiG database, while negative samples can be generated by randomly segmenting bacterial genomes, excluding sequences similar to known BGCs.  Protein sequences can be obtained directly from genome datasets or annotated from genome sequences. For Pfam-level prediction, Pfams are first identified along the protein sequences using bioinformatics tools, and embeddings for each Pfam are generated using Pfam2vec. For amino acid-level prediction, pre-trained protein language models such as ProtBert-BFD embeddings \citep{Elnaggar2022ProtTrans}  are employed to embed the segmented amino acid sequences. Once these embeddings are obtained, deep learning models are applied to assign scores, indicating the probability that each embedding corresponds to a BGC.

\section{Public knowledge integration in microbiome studies with LLMs}

Due largely to the rapid development and growth of metagenomics research in the last two decades, it is well established that the human microbiome is associated with overall human-host health. Many of the findings that link the gut microbiome to complex diseases, such as IBD and Crohn's Disease, can be found within individual scientific publications. Manual aggregation of these results, available in the public domain, into an organized and searchable repository would be time-prohibitive and limited to only a small subject of microbes and diseases \citep{badal2019challenges}. Such knowledge bases can be used for downstream analysis and discovery. NLP and test mining approaches can be used to automate this process.

Automated extraction of microbiome-disease associations from scientific text requires three steps. First is to identify the disease and microbe(s) mentioned in the text. This is known as entity extraction, where the entity is either the disease or microbe. Well-established algorithms such as Named entity Recognizers (NERs) and linguistic taggers can be used for this process. The second step is relationship extraction which aims to establish the existence of a relationship between a pair of entities (i.e., microbe-disease pair). The final step is to refine the categorization of identified relationships into positive or negative associations. Several statistical models have been developed for relationship extraction. While each step requires the use of NLP algorithms, the integration of deep learning and LMMs into steps two and three are of recent particular interest.

An early example of using deep learning in relationship extraction comes from \cite{wu2021mining}. In this work, the authors apply a pretrained BERE model to identify microbe-disease associations. BERE is a deep learning model initially developed for extracting drug-related associations \citep{hong2020novel}. The model is pretrained using a biomedical corpus. The model converts the text into vector representation using word embeddings with sentences represented as 200-dimensional concatenations. Then the recurrent neural network encodes short- and long-range dependencies, as well as semantic features using gated recurrent units (GRUs). Finally, a classifier performs prediction. The prediction task has four possible labels: positive in which the microbe's presence will increase when disease occurs, negative in which the microbe's presence will decrease when the disease occurs, relate when the microbe-disease pair occurs together but the relationship cannot be determined, and NA when there is no relationship description in the text. The model requires a large amount of training data. Although, the gold standard of manual curation is difficult and costly. The authors implement a transfer learning silver standard corpus, learned with automated tools but potentially with error, first and then fine tune with the gold standard manually curated corpus. This transfer learning approach results in a reduction in the error rate.

Deep learning models like the one just described have been recently refined to use LLMs like GPT-3 and BERT \citep{karkera2023leveraging}. The principal advantage of using LLMs in this setting is that they reduce the requirement for large amounts of training data, given that they are already pretrained with large amounts of text. the setting where no fine-tuning or training data is used is known as zero-shot learning. \cite{karkera2023leveraging} uses the same positive, negative, relate, and NA labels as \cite{wu2021mining} with their LLMs and find that zero- and few-shot learners do not perform very well, particularly with the NA label. Thus indicating that out-of-the-box implementation of LLMs for identifying microbe-disease associations is limited. The performance of generative (e.g., GPT-3) and discriminative (e.g., BERT) models improve with fine-tuning. The amount of improvement is strongly dependent on the quality of training data.


\section{Discussion}
The recent development of deep learning methods, and large language models in particular, has led to many novel applications that address significant challenges in microbiome and metagenomic research. In this paper, we have reviewed the latest applications of these methods in microbial function analysis, including the identification of biosynthetic gene clusters in bacterial genomes, annotation of virome genomes, and prediction of virus-bacteria interactions. We have also explored the use of generic LLMs, such as ChatGPT, for extracting microbe-disease associations from public knowledge.

While significant strides have been made in analyzing microbiomes using metagenomic data, integrating multi-omics datasets (e.g., transcriptomics, proteomics, metabolomics) remains a crucial area for future research. Deep learning models and large language models capable of seamlessly integrating these diverse data types could provide a more holistic understanding of microbial functions and interactions, leading to more accurate predictions and novel discoveries in microbial ecology.

To further advance this promising research area, it is essential to focus on both the collection and annotation of datasets from multiple sources and the development of new deep-learning architectures specifically tailored for microbiome and metagenomic analysis. The integration of diverse datasets—ranging from genomic sequences to environmental metadata—will provide a more comprehensive understanding of microbial communities and their interactions. However, this requires meticulous data curation, standardization, and the creation of large, well-annotated datasets that can serve as benchmarks for training and evaluating deep learning models.

On the architectural front, there is a need to design models that can handle the unique challenges posed by microbiome and metagenomic data, such as high dimensionality, sparsity, and complex relationships between microbial species. Innovations in model architectures, such as graph neural networks, attention mechanisms, and hierarchical models, could play a crucial role in capturing the intricate dependencies within the data. Moreover, these models should be adaptable to the evolving nature of the datasets, allowing for continuous learning and refinement as new data becomes available.

\bibliography{sn-bibliography}

\end{document}